\documentclass[fleqn,twoside]{article}
\usepackage{espcrc2}
\usepackage{graphicx}
\title{
{\vspace{-1.2em} \parbox{\hsize}{\hbox to \hsize
{\hss  \normalsize TRINLAT-02/02}}} \\
QCD on $2+2$ anisotropic lattices}
\author{Giuseppe~Burgio\address[TCD]{TrinLat Collaboration \\ 
        School of Mathematics, Trinity College, Dublin 2, Ireland}
        \thanks{Talk given by Giuseppe~Burgio},
        Alessandra~Feo\addressmark[TCD],
        Mike~J.~Peardon\addressmark[TCD]
        and
        Sin\'ead~M.~Ryan\addressmark[TCD]}
       
\begin{document}
\begin{abstract}
We discuss the implementation of QCD on $2+2$ anisotropic lattices. 
Technical details regarding the choice of the action as well as perturbative 
and non-perturbative improvement are analyzed. The physical applications 
of the program are presented.
\end{abstract}

\maketitle

\section{Introduction: Why $2+2$?}
\vskip -0.2cm

QCD on $3+1$ anisotropic lattices \cite{ratpack} has recently proved to be a successful tool for 
non-perturbative 
investigation of various interesting physical phenomema (glueball spectrum, $b \bar{b}$ states, 
gluon strings, spectral density). 
Although the method is successful for processes where
the hadronic final state is static in the center of mass frame, it is less 
suited to processes where the hadron momentum in the final state is non-negligible compared to its 
mass. In such cases independent fine space directions for each momentum
degree of freedom are needed to describe the system, increasing dramatically the computational cost and
the difficulties linked with tuning the parameters.
However, there exist some interesting physical phenomena, e.g. $B\!\rightarrow\!K^{*}\, \gamma$, $B\!
\rightarrow\!\pi\, l\, \nu$ and $B\!\rightarrow\!\rho\,l\,\nu$, where the final hadron 
state is either emitted only or strongly peaked at a momentum roughly equal to its mass.
Although such systems have been investigated in the $3+1$ case \cite{shi0}, 
a $2+2$ lattice discretization 
could allow some improvement in the comparison
with experimental data, keeping the ``technical'' costs at bay. Study of other high-momentum form 
factors could also benefit from such a program.

\section{Setting the Framework: Pure $SU(N_c)$}
\vskip -0.2cm

Starting from the simple $1 \times 1$ plaquette action
$$
S_{\rm gluon} = \beta \sum_{x,\mu,\nu} \left( 1 - \frac{1}{2 N_c} {c_{\mu \nu}} \, {\rm Re} \, 
{\rm Tr} \, {P_{\mu \nu}}(x)\right),
$$
where the coefficients $c_{\mu \nu}$ must be fixed, either non-per\-tur\-bati\-vely \cite{Alford:2000an} 
or perturbatively \cite{Drummond:2002yg}, 
imposing the recovery of Lorentz invariance in the continuum limit. In the non-perturbative case 
the coefficients are fixed to restore rotational invariance for, say, the static interquark 
potential. In lattice perturbation theory the tree level terms are fixed by the naive recovery of the 
Yang-Mills lagrangian, while higher order corrections in powers of $g^2$ will be fixed by the symmetries 
of the renormalized propagator. The $c_{\mu \nu}$ in lattice perturbation theory are defined as 
\begin{equation}  
              c_{\mu\nu}= \left\{ \begin{array}{lll}
              \xi^2(1+c_{f}^{(1)} g^2 +O(g^4)) & \, \mbox{f-f};\\
              1+\eta^{(1)} g^2 +O(g^4)         & \, \mbox{c-f};\\
              \frac{1}{\xi^2}(1+c_{c}^{(1)} g^2 +O(g^4))& \, \mbox{c-c}\end{array} \right.
\label{cs}
\end{equation}
where $\xi = a_c/a_f$ is the asymmetry ratio while $c_{f}^{(1)}$, $\eta^{(1)}$ and $c_{c}^{(1)}$
are the coefficients for the fine-fine, fine-coarse and coarse-coarse directions.

Applying lattice perturbative calculations also permits a
study of the renormalizability of the theory, important not only for its own sake but also for the 
extraction of physical matrix elements in the decay processes.
The hypercubic physical volume $V$ is fixed and an 
ultraviolet cutoff, bigger in the $z, \,t$ directions
$\lambda_f = \xi \lambda_c$, is imposed such that $V=L_{c}^2 L_{f}^2 \xi^2 a^4$ where $L_f=\xi L_c$ are
the number of sites in the fine and coarse directions. The physical limits are then:
$L_{c,f}\!\rightarrow\!\infty$ and $a\!\rightarrow\!0$ keeping $\xi$ fixed. The aim is to make contact with 
``symmetric'' physics, so the procedure is to relate the anisotropic lattice regularization to an
asymmetric continuum regularization and from there to a standard continuum regularization.
The procedure is general and holds for any anisotropy, including 3+1.

\section{Perturbative vs. Non-Perturbative}
\vskip -0.2cm

To perform a perturbative renormalization of the theory the different pieces of 
asymmetric lattice perturbation theory must be established. Following the 
symmetric case, the action for the gauge fixing, Faddeev-Popov and measure terms are determined.
It will be a general property of Feynman rules not to carry, at tree level, explicit dependence from
the asymmetry, where a continuum analogue exists. The only difference resides
in the Brillouin zones. On the other hand, at the 1-loop level Lorentz invariance will be broken. 
Corrections given by the 1-loop $c_{\mu\nu}$ in Eq.(\ref{cs}) are needed to restore it.
Moreover, a renormalization procedure must be chosen. In analogy with
the symmetric case, PBC are kept and BPHZ with a massive gluon propagator as
an intermediate infrared regulator is chosen. Contact with continuum theory is straightforward.

Feynman rules are better calculated in terms of dimensionless quantities
\begin{eqnarray}
\phi_\mu(n)&=& \phi^b_\mu(n) T^b\rightarrow a [\xi] \, g A^b_{\mu}(x) T^b \nonumber\\
\sum_n & \rightarrow & \frac{1}{\xi^2 a^4} \sum_x
\end{eqnarray}
where the factor $[\xi]$ will be included according to the value of $\mu$. 
In this way the pure gauge term reads in the continuum limit as
\begin{eqnarray}
S &=& \frac{1}{4} \sum_n c_{\mu\nu}{\hat{F}^b_{\mu\nu}}(n) {\hat{F}^b_{\mu\nu}}(n)+{\cal O}\nonumber\\
&=& \frac{1}{4}\sum_{x} {F^b_{\mu\nu}}(x) {F^b_{\mu\nu}}(x)+O(a^2)
\end{eqnarray}
where $\cal O$ is an irrelevant operator and
\begin{eqnarray*}
\hat{F}_{\mu\nu} &=&  \hat{\partial}^R_{\mu} \phi_{\nu}-\hat{\partial}^R_{\nu} 
\phi_{\mu}+ig[\phi_{\mu}, \phi_{\nu}]\\
F_{\mu\nu}&=&\partial^R_{\mu} A_{\nu}-\partial^R_{\nu} A_{\mu}+ig[A_{\mu}, A_{\nu}]\\
\partial^{R}_{\mu} \varphi(x)&=&\frac{\varphi(x+a [\xi] \hat{\mu})- \varphi(x)}{a[\xi]}\\
\hat{\partial}^{R}_{\mu} \varphi(n)&=& \varphi(n+\hat{\mu})- \varphi(n) \, .
\end{eqnarray*}
The gauge-fixing term is
\begin{eqnarray}
S_{\rm gf}&=&\frac{1}{\alpha} \sum_x {\rm Tr}\left(\partial^{L}_{\mu} A_{\mu} 
\right)^2 \\ 
&=&\frac{1}{\tilde{\alpha}}\sum_n {\rm Tr}\left({c^{(k)}_\mu \hat{\partial}^{L}_{\mu} \phi_\mu(n)}\right)^2\,,
\nonumber
\end{eqnarray}
and the tree-level propagator reads
\begin{eqnarray}
\Pi^{(0)}_{\mu\nu}={{\delta^{ab}}\over{\hat{k}^2}}\left(\delta_{\mu\nu}-(1-\alpha)\frac{\hat{k}_{\mu} \hat{k}_{\nu}}
{\hat{k}^2}\right)\,,
\end{eqnarray}
where the only explicit dependence on the asymmetry is carried by
$$\hat{k}_{\mu}=\frac{2}{a[\xi]}\sin\bigg(\frac{a[\xi] k_{\mu}}{2}\bigg)\,. $$
Similarly for $S_{\rm meas}$ and $S_{\rm FP}$:
$$
S_{\rm meas}= - \frac{1}{2} \sum_{n,\mu} {\rm Tr} \log \left({2 \left(1-\cos \tilde{\phi}_{\mu}(n)
\right)}\over{\tilde{\phi}^2_{\mu}(n)}\right)
$$
$$
S_{\rm FP}=-\sum_{n,\mu} \bar{\hat{c}}^a (n) c^{(k)}_{\mu}\hat{\partial}^L_{\mu}\hat{D}^{ab}_{\mu}
(\phi)\hat{c}^b(n)\,,
$$
where ($t^a$ are the adjoint generators):
$$
\tilde{\phi}_{\mu}(n)=\phi^a_\mu(n) t^a, \;\hat{D}_{\mu}(\phi)= M^{-1}(\tilde{\phi}_{\mu}) 
\hat{\partial}^R_{\mu}+i \tilde{\phi}_{\mu} \,.
$$
As is clear from above, the $g^2$ measure and the 4-gluon and 2-ghost-2-gluon vertices 
will carry an explicit dependence on the asymmetry, corresponding to 
non Lorentz 1-loop corrections to $\Pi_{\mu\nu}$. We refer to a forthcoming paper for
the details of the vertices \cite{ours}.
The resulting 1-loop graphs will be polynomials in $\xi, 1/\xi$ and functions of the external momenta 
times integrals
$$B(n,n_a,n_b)= \int_{-\pi}^{\pi}\frac{d^4 k}{(2 \pi)^4}\;\frac{\sin^{2 n_a}\frac{k_{a}}{2} 
\sin^{2 n_b}\frac{k_{b}}{2}}{D(\xi^2,m^2)^n}\,,
$$
$$D(\xi^2,m^2)=\sum_{\mu\in{\rm f}}\sin^{2}\frac{k_{\mu}}{2}+\frac{1}{\xi^2}\sum_{\mu\in{\rm c}}\sin^{2}\frac{k_{\mu}}{2}+m^2\,,
$$
which can be simplified by using relations like
$$
B_f(n,1)=\frac{1}{2} B(n-1) - \frac{m^2}{2}B(n)-\frac{1}{\xi^2}B_c(n,1)\,,
$$
\begin{eqnarray*}
\frac{\partial}{\partial \xi^2}B_c(n,m)&=&\frac{n}{\xi^4}\left[B_c(n+1,m+1)\right.\\
                                       &+&\left.B_c(n+1,m,1)\right]  \, .
\end{eqnarray*}
The basic integrals are straightforward to evaluate with high numerical precision as a function of $\xi$,
e.g. $B(1)=0.15493339, 0.4638551, O(\log\xi^2)$ for $\xi=1, 6, \infty$.
Tadpole improvement is useful to ensure convergence of the perturbative expansion and 
is implemented by setting
$$  c_{\mu\nu}\rightarrow \left\{ \begin{array}{lll}
              c_{{\rm f}{\rm f}} & \mbox{f-f};\\
              c_{{\rm c}{\rm f}}/u^2 & \mbox{c-f};\\
              c_{{\rm c}{\rm c}}/u^4 & \mbox{c-c}\end{array} \right.\;u= 
\frac{1}{N_c}<{\rm Re} \, {\rm Tr} \, P_{{\rm c}{\rm c}}>$$ 
Results are very encouraging. The data for the potential with tree level 
coefficient, including tadpole improvements, already show small breaking of 
Lorentz invariance.
\begin{figure}[htb]
\vskip -0.3cm
\includegraphics[scale=0.4]{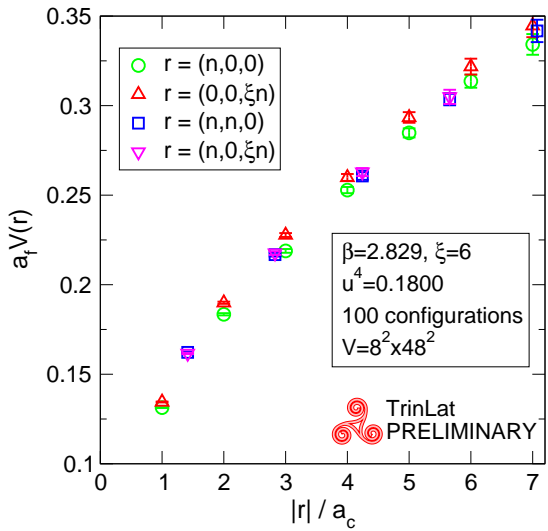}
\vskip -0.8cm
\caption{Static potential for $\xi = 6$, with tree level $c_{\mu\nu}$ in the action.}
\label{fig1}
\vskip -0.7cm
\end{figure}
Preliminary results show that 1-loop corrections are $O(10^{-2})$ 
after tadpole subtraction, confirming that analytic control is very important.

\section{Including quarks}
\vskip -0.2cm

When including fermions, avoiding $O(a_c)$ discretization errors is 
the main problem when dealing with 
asymmetric discretizations \cite{shi}. To circunvent this we choose a formulation which includes up to
$4^{\rm th}$ derivative terms \cite{hw}. The fermion action reads
\begin{eqnarray*}
S_q &=& \bar{\psi} \left( m + \sum_{c} (\gamma_c \Delta^{(1,{\rm i})}_c 
               + s a_c^3 \Delta^{(4)}_c)\right.\\
              &+& \left.\sum_{f} (\gamma_f \Delta^{(1)}_f + \frac{ra_f}{2} \Delta^{(2)}_f)
    \right)\psi
\end{eqnarray*}
where the regular Wilson discretization is used for the fine axis while up to $4^{\rm th}$ order corrections
are included in the coarse directions \cite{hw}.

\section{Outlook \& Developments}
\vskip -0.2cm

We have shown how 2+2 lattice QCD can be relevant to useful physics and feasible,
although parameter tuning is necessary. Our formulation of lattice perturbation theory
gives fine control over renormalization and a framework
for the calculation of matrix elements. Moreover, the quark lagrangian can be tuned using the same techniques,
allowing interesting prospectives for the method \cite{ours}.

\section*{Acknowledgements}
\vskip -0.2cm

This work was partially funded by the Enterprise-Ireland grants SC/2001/306 and 307.

\end{document}